\documentclass[onecolumn, showpacs, preprintnumbers, nofootinbib, aps]{revtex4}
\usepackage{graphicx}

\newcommand{\bi}{\bibitem}
\newcommand{\be}{\begin{eqnarray}}
\newcommand{\ee}{\end{eqnarray}}
\newcommand{\rar}{\rightarrow}

\begin{document}

\title{Apparent shape of super--spinning black holes}

\author{Cosimo~Bambi$^{\rm 1}$}
\email[E-mail: ]{cosimo.bambi@ipmu.jp}
\author{Katherine~Freese$^{\rm 2}$}
\email[E-mail: ]{ktfreese@umich.edu}

\affiliation{$^{\rm 1}$IPMU, University of Tokyo, 
Kashiwa, Chiba 277-8568, Japan \\
$^{\rm 2}$MCTP, University of Michigan, 
Ann Arbor, Michigan 48109, USA}

\date{\today}

\preprint{IPMU08-0095}

\begin{abstract}
We consider the possibility that astrophysical Black Holes (BHs) 
can violate the Kerr bound; i.e., they can have angular momentum 
greater than BH mass, $J > M$. We discuss implications on the BH
apparent shape. Even if the bound is violated by a small amount, 
the shadow cast by the BH changes significantly (it is $\sim$ an order
of magnitude smaller) from the case with 
$J \le M$ and can be used as a clear observational signature in 
the search for super--spinning BHs. We discuss briefly recent 
observations in the mm range of the super--massive BH at the center 
of the Galaxy, speculating on the possibility that it might violate 
the Kerr bound.
\end{abstract}


\maketitle

\section{Introduction}

Black Holes (BHs) are quite strange objects, which are devoid of a 
true internal structure and are completely defined by a few  
parameters~\cite{c-book, fn-book}. In the case of BHs which we may 
possibly find in our Universe, the number of these parameters reduces 
to three: the mass $M$, the charge $Q$, and the spin $J$. In this 
paper we pay close attention to the possible spin of the BH while we set 
$Q=0$. At present good BH candidates~\cite{narayan} include 
super--massive objects ($10^5 - 10^9$~$M_\odot$) at the center of 
galaxies and stellar mass objects ($5 - 20$~$M_\odot$) in X--ray 
binary systems. In both cases, we can infer their mass from dynamical 
arguments, studying the Newtonian orbital motion of stars or gas 
around them~\footnote{It is also possible that a third category of 
BHs exists, intermediate mass BHs ($10^3 - 10^4$~$M_\odot$), 
but so far we do not have much information
about them and, in particular, there are no clear measurements of 
their mass~\cite{cmiller}.}. A challenge today is to measure the 
spin of these objects as may be experimentally feasible in the near 
future. Here the difficulty is that we need to probe the spacetime 
close to the horizon, because spin effects are absent in Newtonian 
gravity and are suppressed at small velocities and large distances.

An important feature is that BHs are expected to respect the Kerr 
bound $J \le M$. This is just the condition to have a horizon. If the 
Kerr bound were violated, instead of a rotating BH we would have a 
naked singularity. To see this we can examine the 4 dimensional 
Kerr--Neumann or Reissner--N{\"o}rdstrom solutions. The position of 
the horizon is given by the expression~\cite{mtw,lppt}
\be
R_{H} = M + \sqrt{M^2 - Q^2 - J^2} \, ,
\label{r-hor}
\ee 
where $Q$ and $J$ are, respectively, electric charge and angular
momentum of BH. It is clear that in 4D spacetime the horizon 
cannot be formed if
\be
M < \sqrt{Q^2 + J^2} \, .
\label{mass-limit}
\ee
In the absence of a horizon, there would be naked singularities
which are not allowed. Indeed if condition~(\ref{mass-limit}) is 
fulfilled, the Kerr--Newmann metric makes it possible to reach the 
physical singularity at $r=0$ from some large $r$ in finite time 
without crossing any horizon. One could thus consider closed time--like 
curves and violate causality (see e.g. section~66c of~\cite{c-book} 
or ref.~\cite{carter}). For this reason, usually some kind of cosmic 
censorship is assumed and naked singularities are forbidden~\cite{penrose}.

However, it seems reasonable that the singularity at the center of 
BHs has actually no physical meaning and  it is just the symptom of 
the breakdown of classical General Relativity. First, it is difficult 
to believe that all the matter can collapse into an infinitesimal 
volume. Second, this is usually the kind of pathology which is expected 
to be solved at the quantum level. On these general grounds, one is 
tempted to argue that actually there is no singularity at the center 
and that the Kerr bound may be violated. In particular~\cite{horava} 
discusses possible origins of the breach of the Kerr bound in string 
theory.

One more comment is in order here. Since for $J > M$ there is
no horizon, Robinson's theorem does not hold~\cite{robinson}
and thus (at least in classical General Relativity) 
the super--rotating object might not be described by the Kerr metric.
In quantum gravity we simply do not know what happens.

The most promising approach to measure the spin of BHs is often 
believed to be the study of emission lines (notably the fluorescent 
iron K$\alpha$ at 6.4~keV), where $J$ may be deduced by fitting the 
shape of the line~\cite{reynolds}. The method has some weak points. 
In particular, one has to assume some emissivity function (usually 
modeled as a power law in the radius) and that there is no emission 
inside the Innermost Stable Circular Orbit (ISCO). Relaxing these 
assumptions, one can find quite different results~\cite{reynolds2}. 
Another common approach is the X--ray continuum fitting method, 
which is also based on the fact that the inner edge of the accretion 
disk is presumably the ISCO~\cite{shafee1, shafee2, liu}. Here we 
need to know the distance of the BH candidate, its mass, and the disk 
inclination angle, but in a few cases there are sufficiently reliable 
estimates of these quantities. For our purposes, if we want to test 
the Kerr bound, both approaches do not look suitable, because 
there is no clear difference between a Kerr BH near extremality and 
one which violates the Kerr bound by a small amount: the radius of 
the ISCO is a continuous function of $J$, while we would like to 
observe some physical quantity which is discontinuous at $J = M$.

In this work we study the apparent shape of a super--spinning BH and 
we claim that the observation of its shadow could be used to test 
the Kerr bound. The shadow of the BH is the non--illuminated area 
seen by an observer if the BH is in front of a planar light source. 
In realistic situations, it is usually unlikely to have a bright 
source of this kind. Nevertheless, something very similar to the 
shadow can be observed if the BH is surrounded by an emitting medium 
(typically the accreting gas) which is optically thin (and this is 
always possible for enough high frequencies). Here an arbitrarily 
small violation of the Kerr bound makes the horizon disappear and 
changes significantly the apparent shape of the BH: now only the 
photons reaching the center of the BH are lost, while all the others, 
with turning points at finite distances from the center (or at 
distances larger than some scale coming from new physics), are not 
captured and can therefore come back to infinity.

\section{Kerr black holes}

In this section we briefly review the study of the apparent shape of 
a BH which respects the Kerr bound $J \le M$ (for more details, see 
e.g. section~63 of~\cite{c-book} or refs.~\cite{carter, zakharov, sereno}.
Analogous studies of similar objects can be found in 
refs.~\cite{virb1, virb2, hioki}).  
As described above, the shape of the BH is just the boundary of its 
shadow: if you fire a photon inside the shape, it is swallowed by the 
BH; if outside, the photon is not captured. The geodesics equation for 
the radial coordinate $r$ in the Kerr metric in Boyer and Lindquist 
coordinates for massless particles is
\be
\left(r^2 + J^2 \cos^2\theta\right)^2 
\left(\frac{dr}{d\lambda}\right)^2 = \mathcal{R} \, ,
\ee
where $\theta$ is the polar angle, $\lambda$ is the affine parameter, 
and
\be
\mathcal{R} &=& E^2 r^4 + 
\left(J^2 E^2 - L_z^2 - \mathcal{Q}\right) r^2
+ 2 M \left[\left(J E - L_z\right)^2 + \mathcal{Q}\right] r 
- J^2 \mathcal{Q} \, , \\
\mathcal{Q} &=& p_\theta^2 + \cos^2\theta 
\left(\frac{L_z^2}{\sin^2\theta} - J^2 E^2\right) \, .
\ee
Here $E$, $L_z$, and $\mathcal{Q}$ are constants of motion and are, 
respectively, the energy, the component of the angular momentum parallel 
to the spin of the BH, and the so called Cartan constant. For our 
discussion, it is convenient to introduce the variables $\xi = L_z/E$ 
and $\eta = \mathcal{Q}/E^2$, which are related to the 
``celestial coordinates'' of an observer at infinity by
\be\label{xy-eq}
x &=& \frac{\xi}{\sin\theta_{obs}} \, , \nonumber\\
y &=& \pm \left(\eta + J^2 \cos^2\theta_{obs} 
- \xi^2 \cot^2 \theta_{obs}\right)^{1/2} \, ,
\ee
where $\theta_{obs}$ is the angular coordinate of the observer.

\begin{figure}[t]
\par
\begin{center}
\includegraphics[width=5cm,angle=0]{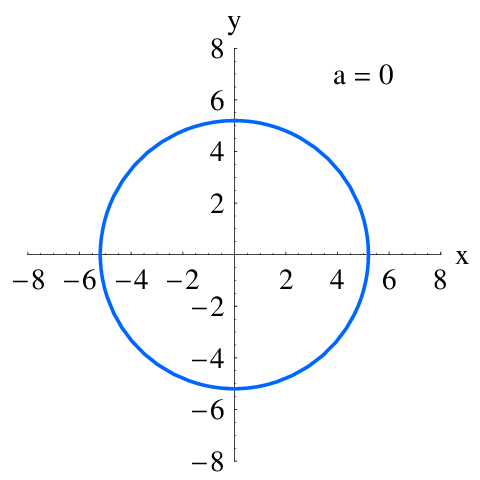} \hspace{.5cm}
\includegraphics[width=5cm,angle=0]{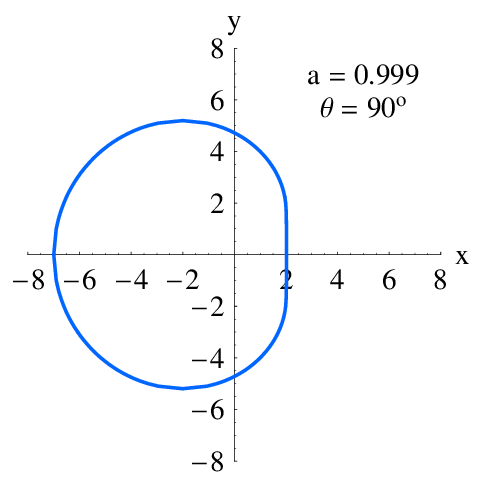} \hspace{.5cm}
\includegraphics[width=5cm,angle=0]{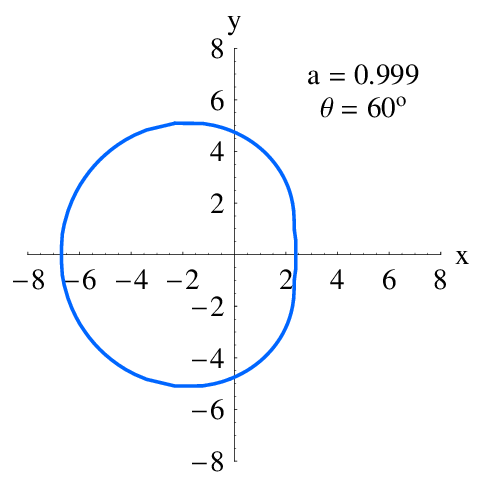}
\end{center}
\par
\vspace{-5mm} 
\caption{{\protect\small 
Apparent shape for Schwarzschild black hole (left panel) and Kerr 
black hole with $a = J/M = 0.999$ (respecting the Kerr bound) for 
an observer with angular coordinate $\theta_{obs} = 90^\circ$ 
(central panel) and $\theta_{obs} = 60^\circ$ (right panel).
The unit of length of the coordinate axes is $M$.}}
\label{fig}
\end{figure}

One can think of an effective potential for the photon, which has
a barrier with a maximum, goes to negative infinity below $r = r_h$, 
where $r_h$ is the horizon, and asymptotes to zero at $r\rightarrow$  
infinity. One can see that there are three kinds of photon orbits:
${\it i)}$ capture orbits, in which the photon arrives from infinity with
energy larger than the barrier of the effective potential
and then crosses into the horizon, ${\it ii)}$ scattering orbits,
in which the photon arrives from infinity with energy less than the
barrier of the effective potential and then comes back to
infinity, and ${\it iii)}$ unstable orbits of constant radius 
(at $r=3M$ for $J=0$, the location of the maximum of the effective 
potential) which 
separate the capture and the scattering orbits~\footnote{In the
simplest case of $J=0$, the effective potential for massless particles 
has a maximum at $r = 3M$, the location of the unstable orbit (there 
is no minimum of this potential). For $J \neq 0$, the picture is 
qualitatively the same, but a little more complex, because the spin
breaks the spherical symmetry of the system. In particular, the 
effective potential is different for particles with angular momentum 
parallel or antiparallel to the BH spin.}.  The apparent shape of the 
BH can be found by looking for the unstable orbits. Every orbit can be 
characterized by the constants of motion $\xi$ and $\eta$, and the set 
of unstable circular orbits $(\xi_c, \eta_c)$ can be used to plot a 
closed curve in the $xy$--plane which represents the apparent shape 
of the BH. The latter is larger than the geometrical shape, because 
the BH bends light rays and thus the actual cross--section is larger 
than the geometrical one. The equations determining the unstable 
orbits of constant radius are
\be\label{r-eq}
\mathcal{R} &=& r^4 + \left(J^2 - \xi^2_c -\eta_c\right) r^2
+ 2 M \left[\left(\xi_c - J\right)^2 + \eta_c\right] r 
- J^2 \eta_c = 0 \, , \nonumber\\
\frac{\partial \mathcal{R}}{\partial r} &=& 
4 r^3 + 2 \left(J^2 - \xi^2_c -\eta_c\right) r 
+ 2 M \left[\left(\xi_c - J\right)^2 + \eta_c\right] = 0 \, .
\ee
In the case of Schwarzschild BH ($J = 0$), the solution is
\be
\eta_c(\xi_c) = 27 \, M^2 - \xi_c^2 \, ,
\ee
so the apparent image of the BH is a circle of radius
$\sqrt{27} \, M \approx 5.20 \, M$ (Fig.~\ref{fig}, left panel). In 
the more general case with $J \neq 0$, one finds
\be
\xi_c &=& \frac{M(r^2 - J^2) - r(r^2 - 2 M r + J^2)}{J (r - M)} \, ,
\nonumber\\
\eta_c &=& \frac{4 J^2 M r^3 - r^4 (r - 3 M)^2}{J^2 (r - M)^2} \, ,
\ee
where $r$ is the radius of the unstable orbit. The apparent shape of
a BH $J = 0.999$~$M$ is reported in Fig.~\ref{fig} for an observer on 
the equatorial plane (central panel) and for one with angular coordinate 
$\theta_{obs} = 60^\circ$ (right panel). The two figures are different,
even if it is not very much evident.

The main feature of the shape of rotating BHs is the asymmetry along
the spin axis, because of the different effective potential for 
photons orbiting around the BH in one or the other direction. The 
radius of the unstable circular orbit is smaller for photons with 
angular momentum parallel to the BH spin and that slightly flattens
the BH shadow on one side. The effect is maximal for the observer on
the equatorial plane, $\theta_{obs} = 90^\circ$. As 
$\theta_{obs} \rar 0^\circ$ (or $180^\circ$), the BH shape reduces 
to a circle of radius $\sqrt{\eta_c(0) + J^2}$, where $\eta_c(0)$ is 
the value of $\eta_c$ for $\xi_c=0$. One can thus find that the 
radius of the circle is a little smaller than $\sqrt{27} \, M$ and 
decreases as the spin increases. For example, when $J = 0.999$~$M$ 
the radius is about 4.83~$M$.

\section{Super--spinning black holes}

If the BH violates the Kerr bound and thus has $J > M$, the picture 
of photon orbits changes. In particular, it is not true that there 
are unstable orbits of constant radius separating the capture and the
scattering trajectories. The apparent shape of the BH can now be 
found looking for the set of points $(\xi_s, \eta_s)$ for which there
is no turning point (and no circular orbits), that is, when 
eq.~(\ref{r-eq}) has no solution for real and positive $r$.  
In classical General Relativity, one generally avoids this 
super-spinning case because (as discussed previously) there is no 
horizon.  Or, one can treat this case by extending the spacetime 
to include regions with negative value of the radial coordinate $r$.  
Then when $J > M$, there are photon orbits with a turning point at 
$r<0$ and thus carry information from ``another universe'' before 
coming back to infinity~\cite{c-book}.

\begin{figure}[t]
\par
\begin{center}
\includegraphics[width=5cm,angle=0]{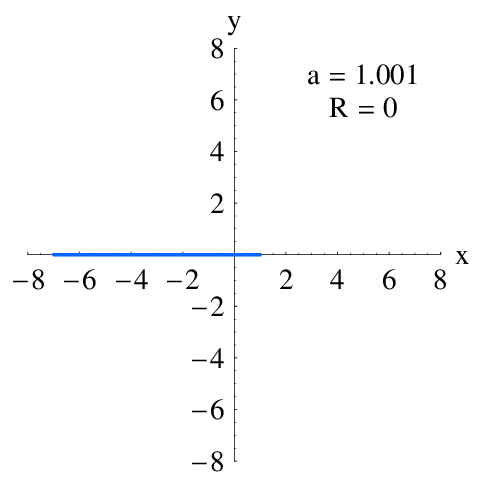} \hspace{.5cm}
\includegraphics[width=5cm,angle=0]{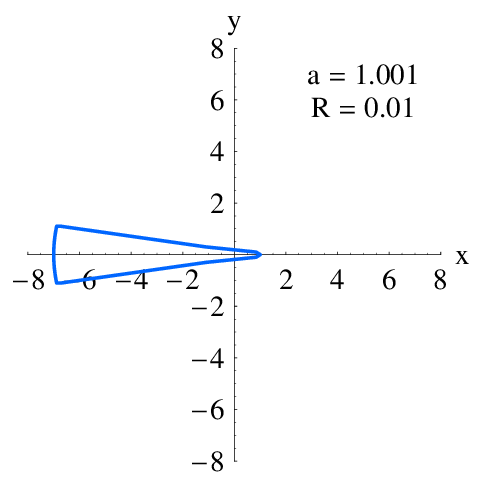} \hspace{.5cm}
\includegraphics[width=5cm,angle=0]{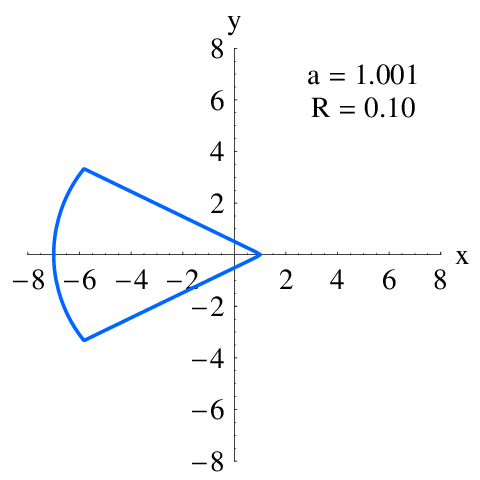}
\end{center}
\par
\vspace{-5mm} 
\caption{{\protect\small 
Apparent shape of a black hole with $a = J/M = 1.001$ for an observer on
the equatorial plane. Here we demand that the distance of the turning 
point of photon orbit from the center is larger than 0 (left panel), 
0.01 (central panel), and 0.10 (right panel).
The unit of length of the coordinate axes is $M$.}} 
\label{fig-ss}
\end{figure}

Here, instead, in a possible extension of GR we believe that it is 
more reasonable to expect that such photons are captured by the 
object replacing the singularity.  Here we require that the turning 
point is at $r > 0$ (or even at $r > R$, where $R$ is some new distance, 
see below) because we are assuming that the region of high curvature
is modified by quantum effects, but we do not know how. For given 
$\eta_s$, one can solve eq.~(\ref{r-eq}) to find $\xi_s$ as a function 
of $r$
\be
\xi_s = \frac{2 J M r \pm \sqrt{4 J^2 M^2 r^2 
- \left[r^4 + (J^2 - \eta_s) r^2 + 2 M (J^2 + \eta_s) r 
- J^2 \eta_s\right](4 M r - r^2)}}{2 M r - r^2} \, .
\ee
For $\eta_s = 0$, there are no solutions for $\xi_s$ in the interval
$(-(6\cosh\chi + \cosh3\chi)M, J)$, where $\cosh3\chi=J/M$, for 
any $r > 0$. On the other hand, for $\eta_s > 0$, $\xi_s$ can have
any value. The apparent image of a BH with $J = 1.001 \, M$ for 
an observer on the equatorial plane is shown in Fig.~\ref{fig-ss},
left panel. Since we are assuming that classical General Relativity 
breaks down (our basic ingredient to consider the possibility of
violation of the Kerr bound), but we do not know what the spacetime
near the former singularity could be, we may make the following 
proposal. One may imagine that quantum gravity effects replace the 
singularity with something larger, say a core of radius $R$; we then 
demand that the radius of the turning point of photon orbits is 
larger than this distance $R$.  We may require that only photons with 
turning point at $r > R$ can really come back to infinity and be 
detected by the observer. In this case, as $R$ increases, the photon 
capture cross--section would also increase.  Fig.~\ref{fig-ss} shows 
the case $R=0$ (left panel),  $R = 0.01 \, M$ (central panel), and 
$R = 0.10 \, M$ (right panel). Even if such a proposal could sound 
crazy in standard General Relativity, it is likely the simplest way
to parametrize new physics.

Quantum gravity effects are presumably important in the region where 
the spacetime curvature approaches the Planck scale; i.e.,  very 
close to the center of the BH. Thus the radius $R$ is likely to be 
very small, i.e., close to 0.  The case $R = 0.01 \, M$ is still 
conservative, since the curvature of the spacetime for astrophysical 
BHs is still tiny (in Planck units) at the distance $0.01 \, M$ from 
the center.  The shape of the BH (the cross section for capture) is 
still very small.  The purpose of considering several values of $R$ 
is just to show some (reasonable?) alternatives. Moreover, we
do not know if a sphere is the best choice for the shape of this
boundary, especially given the extreme rotation of the object. For
example, an oblate spheroid could be a reasonable possibility.
For an oblate spheroid, the apparent shape of the BH would be something in
between  the two spherical cases with $R$ equal, respectively,
to the length of the major and minor axes. However, probably only a
significant deviation from  spherical symmetry could be 
distinguished observationally.

\begin{figure}[t]
\par
\begin{center}
\includegraphics[width=5cm,angle=0]{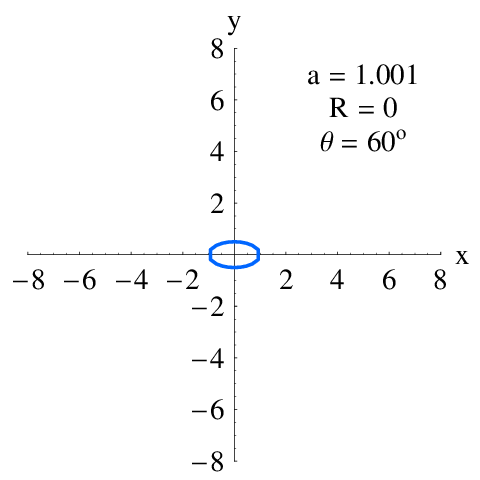} \hspace{.5cm}
\includegraphics[width=5cm,angle=0]{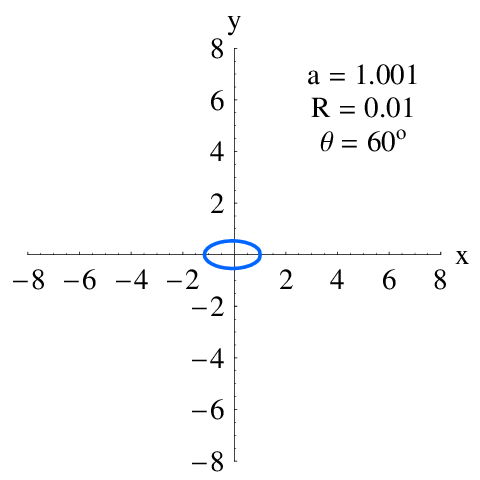} \hspace{.5cm}
\includegraphics[width=5cm,angle=0]{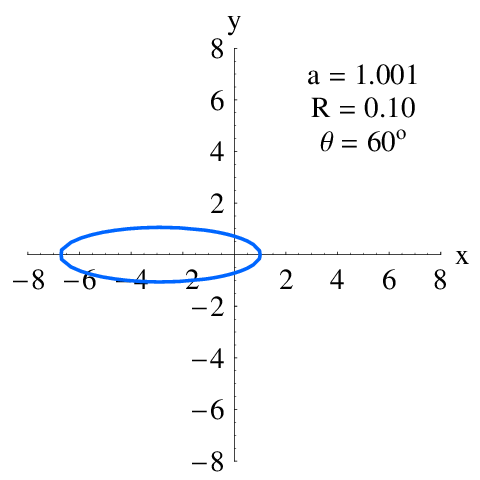}
\end{center}
\par
\vspace{-5mm} 
\caption{{\protect\small 
Apparent shape of a black hole with $a = J/M = 1.001$. Here we demand 
that the distance of the turning point of photon orbit from the center is 
larger than 0 (left panel), 0.01 (central panel), and 0.10 (right panel).
The unit of length of the coordinate axes is $M$. }}
\label{fig-ss-i}
\end{figure}

Let us now study the BH shape for an observer not on the 
equatorial plane. Using eq.~(\ref{xy-eq}), it is easy to see how the 
shape changes. The case $\theta_{obs} = 60^\circ$ is reported in 
Fig.~\ref{fig-ss-i}. The shape in the left panel is for the case of 
photons which can have a turning point arbitrarily close to the origin 
$r = 0$. The photons inside the curves have $\eta_s < 0$ and cannot 
be seen by observers on the equatorial plane (indeed eq.~(\ref{xy-eq}) 
would provide imaginary value of $y$).
The apparent shape of the BH is an ellipse with semiaxis along the 
$x$ direction equal to $J$ and semiaxis along the $y$ direction equal 
to $J|\cos\theta_{obs}|$. For $\theta_{obs} \rar 0^\circ$ (or $180^\circ$), 
the system looks spherically symmetric and the ellipse reduces to a 
circle of radius $J$.

If we demand that only photons with a turning point at radii larger
than some finite value $R$ can come back to infinity, the BH shadow 
for observers with $\theta_{obs} \neq 90^\circ$ is of the kind shown 
in the central and right panels of Fig.~\ref{fig-ss-i}. As expected, 
the shape is larger than the case with $R = 0$, but still much smaller 
than the one of BHs respecting the Kerr bound. For observers with 
angular coordinate $\theta_{obs} = 0^\circ$ or $180^\circ$, the BH
shadow reduces to a circle of radius $\sqrt{\eta_s(0) + J^2}$. For 
$R = 0.01 \, M$ and $R = 0.10 \, M$, one finds respectively 
$1.018 \, M$ and $1.123 \, M$.

\section{Observational consequences}

Quantum gravity effects may resolve BH singularities and in the 
process allow for violations of the Kerr bound, i.e., allow for 
super-spinning BH with $J > M$.  We have shown that a BH violating 
the Kerr bound has no event horizon and has a very different 
apparent shape, i.e. cross section for capturing photons. 
Yet corrections to the spacetime structure are likely to be negligible 
for astrophysical BHs, because the curvature approaches the Planck 
scale only in the very central region. We would like to stress that 
super--spinning BHs cannot be created by spinning up BHs with $J < M$: 
on the basis of the Third Law of BH thermodynamics, there are no 
physical processes capable of transforming a BH with $J < M$ into an 
extremal one in a finite number of steps. So, super--spinning BHs 
must be born as super--spinning BHs. That would violate the cosmic 
censorship conjecture, but there are some known examples which look 
physically reasonable and where it is indeed possible to create a 
naked singularity from the gravitational collapse of matter, see e.g. 
section~5.7.1 of~\cite{fn-book} and references therein.
For example, in $2+1$ dimensions where the study of gravitational
collapse is more tractable, recent results show that, under general
initial conditions, the collapse of a shell with pressure can form
a naked singularity and that, in general, angular momentum does not
prevent the violation of the cosmic censorship conjecture~\cite{mann}.

The possibility of testing the Kerr bound $J \le M$ is particularly
intriguing for two important reasons. First, even though there are not
yet reliable measurements of the BH spin, several indications
suggest that astrophysical BHs typically have high spins. For example, 
in ref.~\cite{reynolds3} the authors find $J/M > 0.93$ for the AGN 
MCG-6-30-15, while the authors of ref.~\cite{miller} suggest 
$J/M > 0.8 - 0.9$ for the stellar mass BH in the X--ray source GX~339-4.
Based on the X--ray continuum fitting method, the lower limit
on $J/M$ of the BH candidate in the X--ray source GSR~1915+105 has
been estimated to be 0.98 in ref.~\cite{shafee2}. 
The second important point is that observations in the mm range are now 
reaching resolutions smaller than the expected angular size of the BH
at the center of the Galaxy~\cite{doeleman}. Another promising candidate
is the super--massive BH in the center of the galaxy M87, which is about 
2000 times more distant, but 1000 times more massive; thus its expected 
angular size is only a factor 0.5 smaller than the one of the BH in 
our own Galaxy.

If a BH is in front of a light planar source, a distant observer sees 
its shadow, a non--illuminated area with boundary equal to the BH
apparent shape.  However, the region blocked by the BH is not likely
to be completely dark.  The BH is likely to be accreting from a disk
which emits radiation itself. Hence, the part of the disk in front of the
BH prevents the BH region from looking completely black; instead, if one
is looking in the direction of the BH, one is likely to see a region
of reduced illumination rather than a completely dark one.
The observer can then see
a less illuminated area which has the same form of the shadow. 
The boundary of this area is not as well defined as the one of the 
shadow:  particles tend to pile up near the last
stable orbits.  Particles sink to the BH because they lose angular 
momentum and thus their barrier decreases. This is not a fast process 
and so matter accumulates around stable orbits and there is the 
possibility of relevant photon emission, which can somehow compensate 
the attenuation due to photon redshift.

The rotation of the accreting matter which emits radiation introduces
an additional source of uncertainty, deforming this darker image in a 
way that Schwarzschild or slow rotating BHs may be interpreted as BHs 
with higher spin value. In the case of super--spinning BHs, basically 
all the emitted photons can reach the observer at infinity, at least 
in principle. Photons which are emitted at distances of order $M$ are 
strongly redshifted if the quantity $J/M$ is slightly larger than one, 
but at the same time they presumably orbit around the BH several times, 
since the orbits are stable, thus increasing the intensity of the flux.

The possibility of observing the shape of the BH at the center of 
the Galaxy was first discussed in~\cite{melia}. The shape should in 
principle be observable at sub--millimeter wavelengths.  Yet, if the 
spin is below the Kerr bound, measurements of the shape will not 
definitively determine the spin. The size of the BH shadow turns out 
to be roughly 10~$M$, regardless of the value of the spin, and even 
if the accretion gas were optically thick and geometrically 
thin~\cite{takahashi}. The measurement of $J$ may instead be achievable 
through more sophisticated multi--wavelength studies of the BH image 
and of its polarization~\cite{broderick}.  Although a precise 
determination of the spin may be difficult, it may be much easier to 
distinguish whether the spin is above or below the Kerr bound.  If the 
BH violates the Kerr bound, the apparent shape of the BH would be 
much smaller than 10~$M$, for any angular coordinate of the observer. 
For example, if $\theta_{obs} = 0^\circ$ or $180^\circ$, the shadow of 
a Kerr BH is a circle of radius in the range $5.20 - 4.83 \, M$; 
whereas a BH slightly in violation of the Kerr bound ($J/M$ a little 
larger than one) has a shadow which is a circle of radius about 1~$M$.

We conclude this section with a speculation on the possible value for 
the spin of the BH at the center of our Galaxy. In ref.~\cite{doeleman},
the authors reported the observation at the wavelength of 1.3~mm
of the radio source Sgr~A$^*$, which is coincident with the position 
of the BH candidate at the level of 10~mas~\cite{ghez}. In fact it 
is not clear whether the radio source is exactly centered on the 
BH~\cite{narayan2} or somewhat shifted away from it~\cite{falcke}. 
Modelling Sgr~A$^*$ as a circular Gaussian brightness distribution, the 
authors of ref.~\cite{doeleman} find that the intrinsic diameter of the 
radio source is $37^{+16}_{-10}$~$\mu$as at $3\sigma$.  However, in 
classical GR, if Sgr~A$^*$ were a spherically symmetric photosphere 
centered on the BH, one would expect a much larger diameter: the minimum 
apparent diameter would range from 10.4~$M$ corresponding to $52$~$\mu$as 
for a non-spinning BH ($J = 0$),  to 9~$M$  corresponding to $45$~$\mu$as 
for a BH spinning at the Kerr bound ($J = M$) and $\theta_{obs} = 90^\circ$.  
Although the current data are not yet capable of absolute confirmation 
of such a measurement, its implications would be interesting. One 
possibility is that the radio source is not perfectly centered at the BH. 
A second possibility is that the radio emission region is indeed a 
photosphere centered on the BH, but the BH violates the Kerr bound and 
thus the emission region may have a small apparent size as discussed in 
this paper, even smaller than 45~$\mu$as. Independent measurements of 
the spin of the BH candidate in the Galactic Center could be possible 
in the near future, for example, by observing signatures of time variable 
structure~\cite{doeleman2}. Here the idea is that the observed 
periodicity is due to hot spots orbiting the BH at a few gravitational 
radii: if that is correct, the fastest one could be associated with the
orbital period at the ISCO, which depends on the BH spin and is much 
shorter for a fast--rotating Kerr BH than for a Schwarzschild BH.

Because of its very low quiescent luminosity in the near IR, it has
been argued that the BH candidate in the Galactic Center cannot be
an object with a hard surface and must have an event 
horizon~\cite{brod-nar1, brod-nar2}. Our proposal may be an 
alternative possibility: an object with neither a solid surface like
a star nor an event horizon like a true BH.

\section{Summary and conclusions}

In classical General Relativity, BH spinning more rapidly than 
the Kerr bound, i.e., spinning with $J > M$, would imply the 
existence of a naked singularity and the violation of causality. 
However, if quantum gravity effects can resolve the singularity, 
causality can be restored and we do not need the Kerr bound. 
Then super--spinning black holes may exist.

In this paper we have discussed how we may observationally identify
a black hole which violates the Kerr bound. The key point is the
absence of the horizon, which leads to a very different apparent 
shape for the black hole. By shape we mean essentially the cross 
section for capturing photons.
If the black hole is surrounded by accreting gas which is optically 
thin, as we believe is the case for the black hole at the center of 
the Galaxy for sub--millimeter wavelengths, we can presumably see 
something similar to the black hole shadow. For the standard $J \le M$, 
the precise measurement of the black hole spin is difficult because 
the image size is always about 10~$M$. On the other hand, the 
observational difference between BH with $J<M$ and with
$J>M$ should be quite dramatic.
The test of the Kerr bound can instead be relatively easy, because 
we have just to be able to distinguish an image of apparent size 
about 10~$M$ (for the case where the Kerr bound is satisfied) 
from one of about 2~$M$(where the Kerr bound is violated). 
A more detailed study of the 
picture is  necessary; in particular, the black hole image has 
to be evaluated in a more realistic framework, assuming some 
astrophysical model for the emitting region surrounding the black 
hole. This will be the subject of another work.

The possible violation of the Kerr bound does not strictly imply the 
breakdown of classical General Relativity, since the theory does not
require $J \le M$, but is surely something which would be unexpected 
in the standard framework and which demands new physics.

\begin{acknowledgments}
We would like to thank David Garfinkle for useful comments.
C.B. was supported by 
World Premier International Research Center Initiative 
(WPI Initiative), MEXT, Japan. 
K.F. was supported in
part by the MCTP and DOE under grant DOE-FG02-95ER40899.  
\end{acknowledgments}

\end{document}